\documentclass[twocolumn,amsmath,amssymb,prl,10pt,nofootinbib,superscriptaddress]{revtex4}
\usepackage{ graphicx, float,amsmath, amsmath, amssymb}
\usepackage[export]{adjustbox}
\newcommand{\qsubrm}[2]{{#1}_{\scriptscriptstyle{\textrm{#2}}}}

\def\be{\begin{equation}}
\def\ee{\end{equation}}
\def\bea{\begin{eqnarray}}
\def\eea{\end{eqnarray}}
\def\bse{\begin{subequations}}
\def\ese{\end{subequations}}

\usepackage[breaklinks, colorlinks,linkcolor=black, citecolor=black,urlcolor = black]{hyperref}
\begin{document}


%
%
\title{Some Conceptual Issues in Loop Quantum Cosmology}

\author{Aur\'elien Barrau\footnote{barrau@lpsc.in2p3.fr} \,and Boris Bolliet\footnote{bolliet@lpsc.in2p3.fr}}%
\address{%
Laboratoire de Physique Subatomique et de Cosmologie, Universit\'e Grenoble-Alpes, CNRS/IN2P3\\
53, avenue des Martyrs, 38026 Grenoble cedex, France
}
%


\begin{abstract}
Loop quantum gravity is a mature theory. To proceed to explicit calculations in cosmology, it is necessary to make assumptions and simplifications based on the symmetries of the cosmological setting. Symmetry reduction is especially critical when dealing with cosmological perturbations. The present article reviews several approaches to the problem of building a consistent formalism that describes the dynamics of perturbations on a quantum spacetime and tries to address their respective strengths and weaknesses. We  also review the main open issues in loop quantum cosmology.
 \end{abstract}

\maketitle

\section{General considerations}
In the strong curvature regime of general relativity (GR), quantum gravity should manifest itself as a resulting repulsive force. In cosmology, the total energy density of the universe is expected not to diverge and to remain smaller than an upper bound that can be guessed to be of the order of $\qsubrm{\rho}{max}\simeq\qsubrm{m}{Pl}^4$. The big bang singularity has to be reconsidered. Loop quantum gravity (LQG) suggests that the big bang is replaced by a bounce \cite{Ashtekar:2006rx}. 
The resolution of the initial singularity is achieved in the sense that the operators corresponding to a complete family of Dirac observables, such as matter density, curvature invariant and anisotropic shears, all remain bounded in the physical Hilbert space. Then, the Hamiltonian constraint can in principle be solved,  numerically if not exactly. 
The usual quantization procedure is the so called $\bar{\mu}$-scheme, which was proven to be exempt of infrared divergences. It uses the physical metric and, {\it e.g.}, a massless scalar field for modeling the energy content of the quantum Universe. The scalar field is also used as an internal time. The relationship between the quantum states, $\chi$, and the classical geometrical data \cite{AshtekarQM} , \textit{i.e.} the metric tensor, has been formulated within two different approaches: (i) in the embedding approach the quantum phase space is embedded into a classical phase space, (ii) in the truncation approach one computes corrections to the classical dynamics by truncating the quantum phase space to low orders of moments only \cite{Bojowald:2010qm}. \\
\indent Note that the Wheeler-deWitt (WDW) theory is based on a continuous geometry, while in loop quantum cosmology (LQC) the quantum geometry is essentially discrete. Hence, the WDW theory is recovered at small spacetime curvature, but in general not for $\lambda\rightarrow0$, where $\lambda$ is the discreteness parameter. Actually, the continuous limit of LQC does not yield a physical theory \cite{AshtekarSinghReview}. \\

A first conceptual issue arises as it has been argued that given the current lack of control on higher-curvature corrections and dynamical inhomogeneities in LQC, evidence for a bounce at high density is, at present, only circumstantial \cite{eucl3}. Indeed, most studies are assuming Gaussian states or other specific forms of coherent or
semi-classical wave functions. This is questionable in the high-density regime. Since strong quantum backreaction and higher-curvature terms (in addition to those coming from holonomies) are implicitly ignored by choosing a coherent state, it is not {\it a priori} fully guaranteed that all evidences for the existence of a bounce are reliable in loop quantum cosmology -- setting aside the issue of signature change that we will review later. \\
\indent We shall now address this objection and show why the bounce is a reliable and generic consequence of LQC.
Most numerical studies have indeed focused on gaussian wave packets. These are parametrized by a volume at which the initial state is peaked, $v^*$, as well as its spread $\sigma_v$ and the initial scalar field momentum, $p_{\varphi_0}$. Three subclasses were investigated numerically \cite{GuptRobustness}: (i) WDW initial states, (ii) gaussians in volume and (iii) rotated WDW states.
Recently,  non sharply peaked states as well as non gaussian states were also considered \cite{GuptNonGaussian}. A numerical scheme called CHIMERA \cite{Diener:2013uka} is being developed since 2013 at Louisiana State University in order to investigate the robustness of predictions in effective LQC. Presently, these numerical simulations all seem to confirm the robustness of the singularity resolution which was proven analytically for arbitrary states \cite{Ashtekar:2007em}. Moreover, further analytical studies showed that fluctuations around any given state are in tight control around the bounce \cite{Corichi:2007am,Corichi:2011rt,Kaminski:2010yz}.\\

The introduction of positive spatial curvature (spherical models) was dealt with in 2006 \cite{AshtekarSpherical} and extended shortly after \cite{Szulc:2006ep}. Since closed models also retain the classical recollapse predicted by GR, one is led to a cyclic cosmological scenario. The status of hyperbolic models ($k=-1$) is less clear but has also been mostly consistently addressed \cite{Vandersloot:2006ws}. In addition, the robustness of the singularity resolution has also been checked for a negative \cite{Bentivegna:2008bg} and a positive cosmological constant \cite{Pawlowski:2011zf}.\\
\indent Ambiguities related to the quantization procedure for closed anisotropic LQC models were addressed \cite{Corichi:2011pg}: in LQC there is a freedom in choosing between closed holonomies around loops (loop quantization) to define curvature, or open holonomies to define connection.  For anisotropic models with non-trivial spatial curvature, loop quantization is not possible,  while for the isotropic spherical FLRW model both quantizations schemes are available but are not equivalent. We refer the reader to the excellent recent review \cite{Agullo:2013dla} (and references therein) for more details about anisotropic models and a clear introduction to the whole framework.\\

The effective equations derived in LQC and applied to the FLRW universe have already led to a large amount of phenomenological studies, making LQC one of the rare quantum gravity model that can be constrained by observations. \\
\indent Measurements of the cosmic microwave background (CMB) anisotropy power spectrum are probing the early universe \cite{PCP2015}. Within the inflationary paradigm, the seeds of the anisotropic features of the CMB light are the quantum fluctuations of the lowest energy state of a scalar field that filled the universe before radiation domination, during the inflationary phase.
The calculation of the primordial power spectrum of these quantum fluctuations at the end of inflation is well known: if quantum fluctuations are generated during inflation, one expects a nearly scale invariant power spectrum slightly red tilted.  After inflation, quantum fluctuations translate into density and pressure perturbations of the cosmological fluids (matter, radiation, dark matter and dark energy) which can be probed directly in the CMB angular  power spectrum \cite{Liddle:1993fq}.
This prediction is in a very good agreement with CMB observations, although a higher level of accuracy is needed in order to conclude about the content of the primordial universe (\textit{i.e.} one or several scalar fields, what exact shape for the potential, etc) and to definitely exclude alternative models. (It should be stressed that the aim of quantum cosmology might be less to suggest an ``alternative model" to the standard paradigm than to provide it with a satisfactory foundation.)
In LQC the past singularity is resolved. Therefore the quantum theory of cosmological perturbation can in principle be extended to the Planck era  or even to the classical prebounce contracting universe \cite{lqc1,Grain:2009kw,Mielczarek:2010bh}. This is the so-called LQC extension of the inflationary scenario. \\

For a flat FLRW universe, the LQC modified Friedmann equation is
\be
H^2=\frac{8\pi G}{3}\rho(1-\frac{\rho}{\rho_{\scriptscriptstyle{\mathrm{B}}}}),
\ee
while the equation expressing the local conservation of energy remains valid, $\dot{\rho}=-3H(\rho+P)$, with $\rho$ and $P$ the density and pressure of the cosmological fluid. For a single scalar field, $\varphi$, they read
$
\rho=\tfrac{1}{2}\dot{\varphi}^2+V(\varphi)$ and $P=\tfrac{1}{2}\dot{\varphi}^2-V(\varphi).
$
If the potential energy dominates over the kinetic energy for a significant amount of time, the fluid has an effective equation of state which is close to `$-1$', allowing the universe to enter a phase of accelerated expansion. \\
\indent Until 2013, the accuracy of CMB measurements was such that the simplest potential for the scalar field, $V(\varphi)=m^2 \varphi^2 /2$, was enough to account for the data.  The LQC community has therefore focused on this simple model. However, the recent results of the Planck mission suggest that the quadratic potential is disfavored at a two sigma confidence level while the Starobinsky potential, $V(\varphi)\propto(1-e^{-\sqrt{16\pi G/3}\varphi})$, seems to be the best alternative. An LQC analysis with this form of potential was performed  recently \cite{Bonga:2015xna}.  The Starobinsky potential is based on a quadratic correction to the GR action\cite{Starobinsky:1980te}.  As a matter of fact, it was shown \cite{Olmo:2008nf} that LQC does not yield such a higher order curvature coupling: the action corresponding to the LQC bounce can only be written in terms of a non-metric theory with an action whose lowest order term beyond the Ricci scalar seems not to be quadratic. This approach however uses the effective modified Friedmann equation as a starting point and deriving the correct action from the full theory remains a challenge to be addressed in the years to come. 


An estimation of the critical density, reached at the bounce, $\rho_{\scriptscriptstyle{\mathrm{B}}}\approx 0.41 m_\mathrm{Pl}^4$, is obtained when the Barbero-Immirzi parameter takes the standard numerical value suggested by black hole entropy calculation, \textit{i.e.} $\gamma\approx 0.2375$. Once the energy density at the bounce is fixed, the background dynamics can be parametrized by a single number: the value of the scalar field at the bounce, $\varphi_\mathrm{B}$. Several questions have to be answered regarding the background before discussing further the dynamics of cosmological perturbations: (i) How likely is inflation to occur? (ii) Can the inflationary phase obtained in LQC last long enough so that the flatness and horizon problems are solved and the amplitude of the CMB angular spectrum explained? (iii) Are the scales probed by the CMB sensitive to the specific dynamics at the bounce?\\
\indent Defining a consensual notion of probability in cosmology is still an open issue \cite{Schiffrin:2012zf}. It can be argued \cite{AS2011} that there is no more ambiguity in LQC since there is in fact a precise time when the probability has to be evaluated, the bounce. Then, a consistent framework for the definition of probability can be built \cite{Craig:2013mga} and even be generalized to the covariant formulation of LQC \cite{Craig:2016iuw}. Using the Liouville measure that corresponds to the Hamiltonian structure of LQC, the following conclusion was reached: the probability for the desired --{\textit{i.e.} in agreement with CMB measurements-- slow roll  not to occur in an LQC solution is less than three parts in a million. Hence a great deal of fine tuning would be necessary to avoid the slow roll inflation that meets the CMB constraints: inflation is an attractor in LQC \cite{Singh:2006im}. The rare dynamical trajectories that fail to meet the observational constraints are those which correspond to an extreme kinetic energy dominated bounce. \\
\indent In parallel to this study, it was suggested \cite{bl} to adopt another viewpoint to the problem of measure: instead of setting initial data at the bounce, why not setting them in the remote past of the pre-bounce contracting branch of the universe and then compute the resulting probability distribution of $\varphi_\mathrm{B}$? It was claimed that in the remote past (\textit{i.e.} when $\rho\ll\rho_{\scriptscriptstyle{\mathrm{B}}}$ and $H<0$) it is the phase of the field, $\delta:=\arctan(\dot{\varphi}/\sqrt{2V(\varphi)})$, that should be taken as the fundamental random variable with a flat probability distribution (the corresponding measure is  preserved as long as $\rho\ll\rho_{\scriptscriptstyle{\mathrm{B}}}$). Quite surprisingly, this assumption translates into a probability distribution for $\varphi_\mathrm{B}$ that is highly peaked around a value that corresponds to a slow-roll inflation of about $140$ e-folds. This would definitely be long enough to match the CMB constraints which require at least sixty e-folds of inflation.\\
\indent In both cases, for initial conditions at the bounce or in the remote past, it seems well established that LQC leads to a long phase of slow-roll inflation shortly after the bounce (preceded by a brief phase of \textit{super inflation}, with $\dot{H}>0$). Except for initial conditions corresponding to a bounce preceded by a phase of deflation \cite{Bolliet:2015bka}, the number of e-fold of superinflation, $\qsubrm{N}{B}$, as well as the total number of e-folds of inflation, $\qsubrm{N}{tot}$, can be expressed in terms of $\qsubrm{\varphi}{B}$ and $\qsubrm{\rho}{B}$ as $\qsubrm{N}{B}=\tfrac{1}{3}\ln\Gamma$, where $\Gamma:= \sqrt{24\pi G\qsubrm{\rho}{B}}/m$, and $\qsubrm{N}{tot}=2\pi G\qsubrm{\varphi}{i}^2-\tfrac{1}{2}$, with
\bea
\qsubrm{\varphi}{i}&=&\qsubrm{\varphi}{B}+\sqrt{\tfrac{2}{3}}\mathrm{Arcsinh}(\Gamma\sqrt{{2}/{\mathrm{W}\left(z\right)}})\label{eq:phii}
\eea
the scalar field at the start of slow-roll inflation, where the argument of the Lambert $\mathrm{W}$-function is $z:=8\Gamma^{2}\exp(\sqrt{48\pi G}{\qsubrm{\varphi}{B}})$, for the quadratic potential. (Similar formulas can in principle be obtained for arbitrary potentials.)
The important conceptual point we want to emphasize here is that the prediction of a long enough phase of inflation, which is a nice result, is actually {\it not} a prediction specific to the detailed structure of LQC. The `$w=-1$' effective equation of state is a strong attractor and the high probability for inflation to occur is only due to the large values of the energy density reached in the vicinity of the bounce. This condition, together with the existence of a scalar field with a reasonable potential, is basically enough to ensure an inflationary era. So LQC does indeed predict inflation as something ``natural" but only in the sense that (in the isotropic setting) most trajectories goes through a high energy density state.\\

To conclude our general comments, we will discuss wether inflation predicted by LQC can have the required duration so that observable scales  in the CMB correspond to scales that were affected by the specific dynamics of the bounce. On dimensional arguments, one might expect that the scales affected by quantum gravity effects have a wavenumber of order $\qsubrm{k}{B}\approx\qsubrm{a}{B}\sqrt{\qsubrm{\rho}{B}}\qsubrm{m}{Pl}^{-1}$, comparable to the radius of curvature at the bounce. (Actually, smaller scales can be also affected in the deformed algebra approach, and larger scales in the dressed metric approach, as we will explain later.) If the duration of inflation is too long, the scales ``sensitive" to the bounce would now be super-Hubble.
Within the inflationary paradigm, CMB experiments are sensitive to scales that exited the Hubble horizon at about sixty e-folds before the end of inflation and correspond to a physical wavenumber ${k}_{\star}\simeq0.002 \mathrm{Mpc}^{-1}$. Combining this information with the estimates of the number of e-folds of inflation in LQC, one is led to conclude that the value of the scalar field at the bounce must belong to a tiny range of values centered around the Planck mass for CMB experiments to be sensitive to LQC effects \cite{Bolliet:2015raa}. In a recent paper \cite{Agullo2015}, this was studied in details, taking into account the degeneracy of the LQC parameters with the CMB parameters, namely the amplitude of the scalar primordial power spectrum and the tensor-to-scalar ratio. 
It was claimed \cite{Ashtekar:2015dja}  that an appropriate choice of the LQC parameters could in principle solve the anomalies observed for the large angular scales in the CMB. Such a choice would also lead to non-trivial predictions for the CMB polarization modes, as the observable scales in the CMB would be scales affected by the bounce (Planckian scales). However, this strategy holds at the price of adding a number of parameters: the Bogoliubov coefficients used to set the initial state of quantum fluctuations at the bounce \cite{Agullo:2014ica}.  In addition, conceptually, it is not clear that the prediction claimed for the polarized CMB is discriminant, because the scalar and polarized spectra are correlated. If the lack of power at large angular scales is due to a statistical fluctuation, it might very well also have imprints in the polarized spectrum too.\\

Finally, we note that the issue of the bounce has recently been investigated in the framework of group field theory \cite{Oriti:2016ueo}. Strikingly, the effective cosmological dynamics, emerging as the hydrodynamics of simple condensate states, leads to a bounce very similar to the one of LQC.

\section{Perturbations}
The background dynamics in LQC is well defined and compatible with data. This is a first success for the model. Going beyond this  basic requirement implies to deal with perturbations in a consistent way. The task is highly non-trivial and different approaches,  that we now briefly review, are being considered. We then adress the question of their advantages and drawbacks.\\

In a series of papers \cite{agullo1,agullo2,agullo3}, a consistent formalism aimed at deriving the dynamics of 
cosmological perturbations propagating in a quantum background was developed. The starting point for the quantization is not the reduced phase space of the strictly homogeneous and isotropic background, $\Gamma_{{\scriptscriptstyle{\mathrm{FLRW}}}}$, but the reduced phase space of the perturbed FLRW space, $\tilde{\Gamma}$. It  encapsulates both the homogeneous and isotropic degrees of freedom \textit{and} 
the inhomogeneous degrees of freedom at first order in perturbation, $\tilde\Gamma=\Gamma_{\scriptscriptstyle{\mathrm{FLRW}}}\times\Gamma_\mathrm{pert}$, so that any quantum state can be written as the tensor product 
$\Psi(\nu,v_{\scriptscriptstyle{\mathrm{S(T)}}},\varphi)=\Psi_{\scriptscriptstyle{\mathrm{FLRW}}}(\nu,\bar\varphi)\otimes\Psi_\mathrm{pert}(v_{\scriptscriptstyle{\mathrm{S}}},v_{\scriptscriptstyle{\mathrm{T}}},\bar\varphi)$ 
with $\nu$ accouting for the homogeneous and isotropic degrees of freedom, and $v_{\scriptscriptstyle{\mathrm{S(T)}}}$ for scalar (tensor) perturbed degrees of freedom. The background quantization is performed using the usual loop quantization techniques for homogeneous and isotropic geometry. The seminal papers 
mostly focused on states that are sharply peaked, as usually studied in quantum cosmology,
though the framework could be applied to any background state in principle. For the quantum background geometry, 
it is possible to define a metric operator,
\begin{equation}
\hat{g}_{\mu\nu}dx^\mu dx^\nu=\hat{H}^{-1}_{\scriptscriptstyle{\mathrm{FLRW}}}\ell^6\hat{a}^6(\bar\varphi)\hat{H}^{-1}_{\scriptscriptstyle{\mathrm{FLRW}}}d\bar\varphi^2-\hat{a}^2d\vec{x}\cdot d\vec{x},
\end{equation}
with $\hat{H}_{\scriptscriptstyle{\mathrm{FLRW}}}=\hbar\sqrt{\Theta_{(\nu)}}$ the Hamiltonian operator of the isotropic and homogeneous 
background, and $\ell^3$ the volume of the considered fiducial cell, while $\Theta_{(\nu)}$ is the difference operator.
The  dynamics of the physical perturbations is  given by the second order part of the total Hamiltonian (still restricted to the square of the first order perturbations) raised as an operator. The action of the total Hamiltonian on the perturbed part of the Hilbert space does depend on the scale factor of the Universe. The quantization procedure for perturbations relies on techniques well understood for a test scalar field evolving on a quantum geometry \cite{Ashtekar:2009mb}. The basic idea is the following. First, one considers the Hamiltonian operators, 
$-i\hbar\partial_{\bar\varphi} \Psi(\nu,v_{\scriptscriptstyle{\mathrm{S(T)}}},\bar\varphi)=\left[\hat{H}_{\scriptscriptstyle{\mathrm{FLRW}}}+\hat{H}_\mathrm{pert}\right]\Psi(\nu,v_{\scriptscriptstyle{\mathrm{S(T)}}},\bar\varphi)$,
and then switches to the interaction picture. Second, the factor ordering of $\hat{H}_\mathrm{pert}$ is chosen to be consistent with the factor 
ordering of the $\hat{g}_{\mu\nu}dx^\mu dx^\nu$ operator. For tensor perturbations (to illustrate the point) the quantum dynamics in the interaction picture reads 
\begin{widetext}
\begin{eqnarray}
\Psi_{\scriptscriptstyle{\mathrm{FLRW}}}\otimes i\hbar\partial_{\bar\varphi}\Psi_\mathrm{pert}=\frac{1}{2}\displaystyle\int\frac{d^3k}{(2\pi)^3}\left\{32\pi G \left[\hat{H}^{-1}_{\scriptscriptstyle{\mathrm{FLRW}}}\Psi_{\scriptscriptstyle{\mathrm{FLRW}}}(\nu,{\bar\varphi})\right]\otimes\left[\left|\hat\pi_{{\scriptscriptstyle{\mathrm{T}}},\vec{k}}\right|^2\Psi_\mathrm{pert}(v_{\scriptscriptstyle{\mathrm{S(T)}}},{\bar\varphi})\right]\right.\nonumber \\
\left.+\frac{k^2}{32\pi G}\left[\hat{H}^{-1/2}_{\scriptscriptstyle{\mathrm{FLRW}}}\hat{a}^4({\bar\varphi})\hat{H}^{-1/2}_{\scriptscriptstyle{\mathrm{FLRW}}}\Psi_{\scriptscriptstyle{\mathrm{FLRW}}}(\nu,{\bar\varphi})\right]\otimes\left[\left|\hat{v}_{{\scriptscriptstyle{\mathrm{T}}},\vec{k}}\right|^2\Psi_\mathrm{pert}(v_{\scriptscriptstyle{\mathrm{S(T)}}},{\bar\varphi})\right]\right\}, \nonumber\\
\end{eqnarray}
\end{widetext}
with $\left(\hat{v}_{{\scriptscriptstyle{\mathrm{T}}},\vec{k}},\hat{\pi}_{{\scriptscriptstyle{\mathrm{T}}},\vec{k}}\right)$ the configuration and momentum 
operators for the perturbation degrees of freedom. Taking the scalar product of the above equation with 
$\Psi_{\scriptscriptstyle{\mathrm{FLRW}}}$ finally leads to the Schr\"odinger equation for the perturbation part of the wave function.
The key point is the 
formal analogy between the quantum dynamics of perturbations evolving on a classical 
background and the quantum dynamics of the perturbations evolving on a fully quantum background. The quantum dynamics can be formally described 
as the dynamics of perturbations in a classical background but with a {\it dressed} metric, {\it i.e.}
\begin{widetext}
\begin{equation}
	i\hbar\partial_{\bar\varphi}\Psi_\mathrm{pert}=\frac{1}{2}\displaystyle\int \frac{d^3k}{(2\pi)^3}\left\{32\pi G(\tilde{p}_{\varphi})^{-1}\left|\hat\pi_{{\scriptscriptstyle{\mathrm{T}}},\vec{k}}\right|^2\Psi_\mathrm{pert}+\frac{k^2}{32\pi G}(\tilde{p}_{\varphi})^{-1}\tilde{a}^4({\bar\varphi})\left|\hat{v}_{{\scriptscriptstyle{\mathrm{T}}},\vec{k}}\right|^2\Psi_\mathrm{pert}\right\},
\end{equation}
\end{widetext}
using the identification
\begin{eqnarray}
	(\tilde{p}_{\varphi})^{-1}=\left<\hat{H}^{-1}_{\scriptscriptstyle{\mathrm{FLRW}}}\right> &~\mathrm{and}~&\tilde{a}^4=\frac{\left<\hat{H}^{-1/2}_{\scriptscriptstyle{\mathrm{FLRW}}}\hat{a}^4({\bar\varphi})\hat{H}^{-1/2}_{\scriptscriptstyle{\mathrm{FLRW}}}\right>}{\left<\hat{H}^{-1}_{\scriptscriptstyle{\mathrm{FLRW}}}\right>}.\nonumber\\
\end{eqnarray}
This dressed metric $\tilde{g}_{\mu\nu}$ is {\it neither} equal to the classical metric nor equal to the metric traced by the sharply peaked background 
state. The final quantization of perturbations can 
be performed with standard techniques of quantum field theory on curved spacetimes  but using the dressed 
metric instead of the classical one. Scalar perturbations have also been calculated in this framework. The  results are that the equations of motion for 
 scalar and tensor perturbations have the same form as in the classical 
case, 
\begin{eqnarray}
&&Q''_k+2\left(\frac{\tilde{a}'}{\tilde{a}}\right)Q'_k+\left(k^2+\tilde{U}\right)Q_k=0, \\
&&h''_k+2\left(\frac{\tilde{a}'}{\tilde{a}}\right)h'_k+k^2h_k=0,
\end{eqnarray}
with $Q_k$ a gauge-invariant variable for scalars, related to the Mukhanov-Sasaki variables via 
$Q_k=(v_{{\scriptscriptstyle{\mathrm{S}}},k}/a)$; $\tilde{U}$ is a dressed potential-like term given by
\begin{equation}
\tilde{U}({\bar\varphi})=\frac{\left<\hat{H}^{-1/2}_{\scriptscriptstyle{\mathrm{FLRW}}}\hat{a}^2({\bar\varphi})\hat{U}({\bar\varphi})\hat{a}^2({\bar\varphi})
\hat{H}^{-1/2}_{\scriptscriptstyle{\mathrm{FLRW}}}\right>}{\left<\hat{H}^{-1/2}_{\scriptscriptstyle{\mathrm{FLRW}}}\hat{a}^4({\bar\varphi})\hat{H}^{-1/2}_{\scriptscriptstyle{\mathrm{FLRW}}}\right>},
\end{equation}
the quantum counterpart of
\begin{equation}
U({\bar\varphi})=a^2\left(fV({\bar\varphi})-2\sqrt{f}\partial_{\bar\varphi} V+\partial^2_{\bar\varphi} V\right),
\end{equation}
with $f:=24\pi G (\dot{\bar\varphi}^2/\rho)$ the fraction of kinetic energy.\\
The power spectrum has been computed in this approach and is nearly scale-invariant, with a slight increase of power at large scales. In a subsequent study, it was understood that due to the freedom one has in selecting the initial state it is also possible to decrease the power at large scale, therefore leading to a better agreement with CMB data. An interesting conceptual point is that the reason why IR modes are affected by quantum gravity effects and not UV ones, as one could naively have expected, is clear and can be summarized as follows  \cite{lqc1}. In LQC the curvature radius is bounded from below and takes its minium non-vanishing value $R_{\scriptscriptstyle{\mathrm{B}}}$ at the bounce. The UV modes, that have wavelengths smaller than $R_{\scriptscriptstyle{\mathrm{B}}}$ do not ``feel" curvature and are in the Bunch-Davies vacuum. While IR modes do `feel' curvature and can be amplified: they might not be in the Bunch-Davies vacuum at the onset of inflation.\\

The main other approach to perturbations in LQC is the ``deformed algebra" approach. The constraints of general relativity form a first class system and this property is key to the consistency of the classical dynamics. It is not {\it a priori} clear whether this delicate consistency remains in effective theories that incorporate quantum corrections. The very notion of spacetime is supposed to emerge from solutions to the fundamental quantum gravity equations. The consistency of the effective quantum-corrected equations has to be ensured before they can be successfully solved.  In some fields of physics, gauge fixing before quantization was shown to be harmless but the case of gravity is much more subtle and intricate than, say, Yang-Mills theories because dynamics is part of the gauge. In the deformed algebra approach this issue is addressed by building the algebra of constraints so that the constraints can be quantized without a classical specifications of gauge or observables. The deformed algebra approach is based on taking care of those gauge issues while embedding GR in a quantum framework.\\
\indent Let us focus here on the holonomy corrections that are well defined and understood. The net effect of these corrections is encoded in the replacement
\begin{equation} \label{replacement}
\bar{k} \rightarrow \mathbb{K}[n] := \frac{\sin(n\bar{\mu} \gamma \bar{k})}{n\bar{\mu}\gamma},
\end{equation}
where $n$ is an unknown integer, $\bar{k}$ is the mean Ashtekar  connection, and $\bar{\mu}$ is the coordinate size of a loop. The quantum-corrected constraints resulting from this substitution are renamed $\mathcal{C}^Q_I$. This replacement, motivated by the fundamental role given to holonomies in LQG, leads to the following algebraic structure:
\begin{equation}
\{ \mathcal{C}^Q_I, \mathcal{C}^Q_J \} = {f^K}_{IJ}(A^j_b,E^a_i) \mathcal{C}^Q_K+
\mathcal{A}_{IJ},
\end{equation}
where the $\mathcal{A}_{IJ}$ terms stand for anomalies and $f^K_{IJ}$ are structure functions (and not anymore constants as in standard Yang-Mills theories). The consistency condition (that is the closure of the algebra) requires $\mathcal{A}_{IJ}=0$. In turn, quite nicely, this condition imposes restrictions on the  form of the quantum corrections, especially when matter is included in the Hamiltonian. Since the result is a modification of the  algebra of constraints of spacetime, as it could be expected from the Hojman-Kuchar-Teitelboim theorem \cite{Hojman:1976vp},
 the quantum structure is not a pseudo-Riemaniann spacetime with a  metric in the usual sense. But it does have a well-defined \emph{canonical} formulation using hypersurface deformations.\\
\indent The conceptual strategy used to determine the algebraic structure can be summarized as follows. The quantum corrected constraints are explicitly calculated for the perturbations up to the desired order. Then, all the Poisson brackets are calculated, therefore exhibiting the anomaly terms. Counter-terms, required to vanish at the classical limit, are finally added to the expressions of constraints to ensure anomaly freedom. The resulting theory is consistent by construction but is also --maybe surprisingly-- uniquely defined: the different unknown integers entering the game (that can be different from unity when one considers other terms than the $\bar{k}^2$  arising from the curvature of the connection)  are all  determined. Furthermore, although the calculations are intricate, the resulting algebra is simple and elegant. It involves a single structure function which encodes, at the effective level, all the quantum modifications:
\begin{equation}
\Omega=1-2\rho/\rho_{\scriptscriptstyle{\mathrm{B}}}. 
\end{equation}
The algebra is then closed in a non-pertubative way. This method has been successful for vector \cite{tom1} and scalar \cite{tom2} 
perturbations. It has aslo been shown that a single algebraic structure can be consistently written for {\it all} perturbations \cite{eucl2}  (this has consequences for tensor modes that were forgotten in first studies), making the whole approach very appealing:
\begin{eqnarray}
\left\{D[M^a],D [N^a]\right\} &=& D[M^b\partial_b N^a-N^b\partial_b M^a], \\
\left\{D[M^a],S^Q[N]\right\} &=& S^Q[M^a\partial_b N-N\partial_a M^a],  \\
\left\{S^Q[M],S^Q[N]\right\} &=& \Omega 
D\left[q^{ab}(M\partial_bN-N\partial_bM)\right],
\end{eqnarray}
where $D$ and $S$ are the diffeomorphism and Hamiltonian constraints, $N$ and $M$ are lapse functions, $N^a$ and $M^a$ are shift functions, and the superscript $Q$ indicates that the constraint is quantum corrected. \\
\indent Beside its elegance, this algebra has a striking feature:  it leads to a signature change close to the bounce. When $\rho<\rho_{\scriptscriptstyle{\mathrm{B}}}/2$ the spacetime structure is Lorentzian but when $\rho>\rho_{\scriptscriptstyle{\mathrm{B}}}/2$, in the vicinity of the bounce, $\Omega$ becomes negative and the spacetime structure becomes Euclidean. This is reminiscent of what is usually postulated in quantum cosmology, mostly for technical reasons (to improve the convergence of path integrals). 
In standard quantum cosmology one usually deals with an amplitude written as
\begin{equation}
<\varphi_2,t_2|\varphi_1,t_1>=\int d[\varphi]e^{I[\varphi]},
\label{eq:qc}
\end{equation}
where $I[\varphi]$ is the action of the field configuration $\varphi(x,t)$, and $d[\varphi]$ is a measure on the space of field configurations. The integrand in \eqref{eq:qc} has a rapidly oscillating phase, and the path integral, in general, does not converge. This is why the time is rotated clockwise by $\pi / 2$ so that $I[\varphi] \rightarrow \tilde{I}[\varphi]:=-iI[\varphi]$. The integrand in the resulting Euclidean path integral is now exponentially damped, and the integral generically converges. Then, one can analytically continue the amplitude in the complex $t$-plane back to real values. Importantly, a quantum field theory machinery has been developed in this framework \cite{Laflamme:1987mx}. Recent reviews on quantum field theory on an Euclidean background can be found in \cite{Strocchi:2013awa,Guerra:2005an}. Although far from being fully understood (but QFT on a Lorentzian curved background is also not completely understood) the framework is basically consistent.
 There are also obvious links with  the Hartle-Hawking proposal \cite{HHP} but the Euclidean phase appears in the LQC model considered here in a fundamentally dynamical way since the Poisson bracket between Hamiltonian constraints varies continuously from a positive to a negative expression. This is the key conceptual point.
 This effect has also been found independently following different approaches within LQC \cite{ed,bp}. In particular, the first of these references relies on substantially different hypotheses (using a model of patches of universe evolving independently in the longitudinal gauge). The fact that it leads to the same result reinforces the credibility of the conclusion, we will come back to this point later. The resulting equation of motion is more complicated than in the ``dressed metric" approach. For tensor modes, it reads
\begin{equation}
		v''_k(\eta)+\left(\Omega k^2-\frac{z_{\scriptscriptstyle{\mathrm{T}}}''}{z_{\scriptscriptstyle{\mathrm{T}}}}\right)v_k(\eta)=0 \label{eom2},
\end{equation}
in conformal time, where  the mode functions $z_{\scriptscriptstyle{\mathrm{T}}}:=({a}/{\sqrt{\Omega}})$  are  related to the amplitude of the tensor modes of the metric perturbation $h_k$ through $v_k=z_{\scriptscriptstyle{\mathrm{T}}} h_k/\sqrt{32\pi G}$. The evolution of the modes is not anymore driven only by the hierarchy between $k^2$ and $|a''/a|$. Due to the $\Omega$-term, the situation is more complicated and several of new phenomena do appear, opening a wide phenomenology. Here, the ratio between the length scale associated with a mode and the curvature radius is not the only important number.\\
\indent By definition there is no time in the Euclidean phase. The very meaning of  ``propagation of a mode" becomes unclear. However, the equation of motion in Fourier space, which reads for scalar modes
\be
\ddot{\mathcal{R}}_k - \left( 3H + 2 m^2 \frac{\bar{\varphi}}{\dot{\bar{\varphi}}} + 
2 \frac{\dot{H}}{H} \right) \dot{\mathcal{R}}_k + \Omega\frac{k^2}{a^2} \mathcal{R}_k = 0,
\label{eomR}
\ee
with
$\mathcal{R} := {v}/{z}$, is mathematically well defined and has a regular solution even if there are singular points in the equation itself. The primordial power spectrum has been calculated and exhibits three regions. This approach implicitly assumes that the change of sign is in fact a kind of tachyonic instability and not a deep change of signature at the fundamental level of the structure of space-time. The case of a real change of signature requires news techniques that are briefly mentioned in the last part of this article whereas  the instability case can be rigorously treated as done in \cite{susanne}. The UV region is then characterized by an exponential growth of the spectrum whose origin is clearly grounded in the Euclidean phase. The intermediate region exhibits oscillations (that would be smeared out by the cosmic evolution). The IR region is mostly scale-invariant for tensor modes and blue for scalar modes. Whether we see the IR, the UV or the intermediate region in the CMB depends on the duration of inflation. If the number of e-folds is higher than the minimum required value, the observational window falls in the UV part, which is incompatible with data. This shows that quantum gravity can indeed be falsified by cosmological experiments, even when it predicts a long enough phase of inflation!\\

What are the conceptual differences between both approaches and the assumptions behind them\footnote{The following arguments come partially from a discussion that took place between I. Agullo, A. Barrau, M. Bojowald, and G. Calcagni.}?  Which one is more reliable? Those questions are not easy to answer. The dressed metric approach is unquestionably more ``quantum" as it addresses the essential question of dealing with a quantum field on a quantum background. In a way,  the fundamental structure is given by the background wave function $\Psi_{\scriptscriptstyle{\mathrm{FLRW}}}$ and the dressed metric itself is only a manner of modeling how perturbations propagate. The background structure is not a smooth Riemannian metric, but a quantum geometry given by $\Psi_{\scriptscriptstyle{\mathrm{FLRW}}}$, which is consistent with the presence of perturbations since  it has been checked that the backreaction they produce is negligible (consistency condition). However one still uses at some point (explicitly or implicitly) a line element $d s^2= g_{\mu\nu} dx^\mu dx^\nu$, with $g_{\mu\nu}$ defined in terms of expectation values. For this to be meaningful, one would still have to demonstrate that $g_{\mu\nu}$, wherever it comes from, changes by standard (classical) coordinate transformations if one changes coordinates (or the gauge). Otherwise, $d s^2= g_{\mu\nu} dx^\mu dx^\nu$ is not coordinate independent and not tensorial, and therefore loses its meaning. The problem is that the deformed algebra approach precisely shows that, in general, the classical transformations do not apply anymore when holonomy corrections are  implemented consistently.\\
\indent The basic question one can therefore ask about the dressed metric approach is whether it really goes beyond quantum field theory on curved spacetimes. If one has a background FLRW metric with a scale factor $a$ and some scalar field $\varphi$, one can quantize any field on it, such as the gauge-invariant tensor modes. The background equations need not solve the classical Friedmann equation because quantum field theory can in principle be done on any Riemannian background, not just on one solving Einstein's equation. So even when one uses expectation values of minisuperspace operators, instead of the classical $a$ and $\varphi$, one might still be within the setting of quantum field theory on a curved spacetime, rather than in quantum gravity. Gravity, even linearized, differs from fields on a background because it determines how the fields transform under coordinate transformations. This is what is meant by having a metric structure as opposed to just a background with fields lying over it. Quantum gravity in the linearized setting should, in general, differ from quantum field theory on curved spacetimes by deriving the existence of a corresponding structure with specific transformations. This is a key point and one must be careful of not implicitly assuming a classical background structure that has no reason to be correct in this framework.\\
\indent There is {\it a priori} an infinite number of dynamical laws, all written with respect to different choices of time coordinates. They are classically equivalent to one another thanks to the symmetries we know GR enjoys, and one is therefore free to pick any one of these choices. In the dressed metric approach, when referring to a background gauge, by deparameterization, or by writing the mode dynamics in terms of coordinate invariant combinations of metric and matter perturbations, one might be implicitly using several time choices. It is only after these steps that a specific dynamics for background variables and perturbations can be obtained. Classically, those results do not depend on which coordinate choices are made, and the procedure is valid. But when one quantizes some (or all) relevant degrees of freedom, the equations are modified by quantum corrections of different kinds, and one is no longer guaranteed that the results do not depend on the choices made (that is, the theory may not be covariant or anomaly-free). What is crucial is the fact that the classical theory enjoys a symmetry which might be used  in order to simplify the quantization procedure. When quantized or modified, the theory does not exhibit this symmetry anymore and the results might be gauge-dependent and therefore meaningless.\\
\indent A possible answer is that  the dressed metric approach does indeed go beyond quantum filed theory on a FLRW spacetime in the sense that perturbations now propagate on a quantum FLRW background. The evolution of perturbations is sensitive to the quantum nature of the background. It is sensitive not only to the fact that the peak of the wave function does not follow the classical evolution, but also to the quantum fluctuations. This, however, does not answer the consistency issue.\\

The other way round, one could wonder if the algebraic structure obtained in the deformed algebra approach is really unique. After all, there is no theorem proving that the values chosen for the free parameters entering the non-linear system under consideration are the only possible ones. And, indeed, they are probably not unique. Assumptions --although quite natural-- need to be made in order to obtain a tractable solution. \\
\indent A more serious criticism is the following. The theory of cosmological perturbations is a truncated theory in which one approximates the exact solutions by throwing away some terms in equations. It is clear that the most obvious way of doing it is to consider Einstein's equations, expand them around a given background, and keep terms up to first order in perturbations. 
It is then possible to recast the same dynamics in the Hamiltonian language. But the evolution of perturbations is not generated by a constraint. The full evolution is not generated by a single Hamiltonian. Rather, background evolution is dictated by a Hamiltonian $H_0$ (the index $i$ in $H_i$ refers to the considered order), and perturbations evolve on the top of this background according to their own Hamiltonian $H_2$. This dynamics differs from the one generated by the single Hamiltonian $H_0+H_2$.  Furthermore, perturbations are constrained by linear constraints such as the ones generated by $H_1$,  but $H_2$ is not constrained to vanish. This suggests that the algebra of second order constraints might not play a fundamental role in the theory.  One could follow a different approach and modify this setting by declaring that the dynamics is generated by the Hamiltonian $H_0+H_2$ and that $H_2$ is also a constraint. The resulting  equations of motion are different from the truncation of Einstein equations mentioned above as the new equations involve some backreaction. However is only involves part of the full backreatcion, which raises a consistency issue.\\ 
\indent This issue could be addressed as follows.  The classical constraint, playing the role of a starting point, reads $H[N]=0$ (there are actually  infinitely many constraints because the lapse function $N$ is free, setting aside the diffeomorphism constraint). If this constraint is expanded, it remains a constraint.  To second order, one can split it in two different terms: one that comes from varying  the background lapse $\bar{N}$ (a single constraint $H_0+H_2$) and one that comes from the variation with respect to $\delta N= N-\bar{N}$ (an infinite number of constraints $H_1$). 
 For unique Hamiltonians $H_0$ and $H_2$ the background lapse has to be fixed and deparameterization can be used. The  constraint $H_1=0$ is solved for the modes.   However, this would not give the correct dynamics $H_2$: for general metric perturbations (no gauge fixing) there is more than one independent scalar degree of freedom left in $H_2$, even after solving the constraint $H_1=0$. It is possible to get rid of non-physical scalars either by fixing the gauge, or by rewriting $H_2$ in terms of gauge-invariant combinations of the scalar perturbations. In both cases, one refers to gauge transformations generated by $H_1$. Treating them as gauge transformations is consistent only if  $H_1$ satisfies a closed first-class algebra. This is why the deformed algebra puts a specific emphasis on this point. The classical closure of this algebra relies on the background equations as well as the mode equations. If the background dynamics is modified by quantum effects, the algebra generated by $H_1$ is no longer guaranteed to close.  It is however true that the dynamics generated by $H_0+H_2$ can, in some background gauge, include a backreaction term and that there may be additional terms which are not taken into account into procedure. The important point is that this makes the system canonical, so the powerful methods of constrained systems can be used. In other words, this term is included for mathematical rather than physical reasons but it is correct to underline that a possible lack of physical consistency on this specific point still need to be addressed.\\
 
On the one hand, it should be emphasized that the main results of the deformed algebra approach have also been found independently following different paths. The holonomy corrections to the effective equations was considered in the longitudinal gauge \cite{WilsonEwing:2011es}. The main result is that the equations of motion for the perturbations agree with the one of the deformed algebra. The algebraic structure is also similar, as the time derivative of the effective scalar constraint gives rise to a cosine multiplying the diffeomorphism constraint.  In another work \cite{WilsonEwing:2012bx}, the quantum theory on a lattice was studied so that long wavelength scalar perturbations in LQC could be accounted for, again using the longitudinal gauge. The commutators are explicitly calculated and taking the classical limit, and then the continuum limit, one recovers the considered algebra. Some support to the deformed algebra approach also comes from recent investigations of a linear redefinition of the constraints (with phase-space dependent coefficients) which can be used to eliminate structure functions, even Abelianizing the more-difficult part of the constraint algebra \cite{Bojowald:2015zha}.\\
\indent On the other hand, an analysis \cite{Gomar:2015oea} of the quantization of cosmological perturbations in which one truncates the action up to second order in perturbations and looks at the whole symplectic system formed by zero-modes plus perturbations was performed. Perturbations are not treated as a test field on a background, as in the dressed metric approach, but backreaction up to the order considered is included. The model is here parametrized using only gauge-invariant quantities: it deals with the construction of approximate solutions of inhomogeneous cosmologies that effectively behave as approximate solutions of a homogeneous and isotropic model with a specific matter content, or even with geometric modifications \cite{Navascues:2014rla,Navascues:2015saa}. These solutions are far from being homogeneous, and the terms accounting for the matter or GR modifications have their origin in the collective behavior of the inhomogeneities. These solutions were constructed in the specific case of the hybrid quantization of the linearly polarized Gowdy model with three-torus topology, as an example of the fact that inhomogeneous quantum degrees of freedom can behave collectively to lead to a homogeneous description.\\

To conclude, let us emphasize that in spite of the conceptual and technical differences between the approaches we have presented, there are {\it universal} LQC features that appear at the phenomenological level (for tensor modes) in the IR and intermediate regions of the primordial power spectrum \cite{Bolliet:2015bka}. This is a pleasing and encouraging result.

\section{Other open conceptual issues}

Ignoring the possible Euclidean phase around the bounce, we first come back to a simple question related to initial conditions. If the bounce is resulting from a causal evolution, there is no reason for initial conditions --neither for the background nor for the perturbations-- to be set at the bounce time. As advocated by the deformed algebra authors \cite{Linsefors:2013cd,Linsefors:2014tna}, this is both causally unjustified and technically irrelevant since the bounce is probably the worst time (strong curvature regime) for setting initial conditions, especially when considering that the causal structure in the remote past of the contracting branch allows one to have a well defined vacuum state in the usual QFT sense. However, the way time flows is not obvious. It is perfectly allowed to assume that, starting from the bounce, time flows in two opposite directions, generating two {\it expanding} branches. In that case, putting initial conditions at the bounce is necessary. In principle both scenarios are distinguishable observationally as some extreme gravitational phenomena occurring in the contracting branch (first hypothesis, with only one time direction) might have footprints in the current universe \cite{Nelson:2011gb}. \\

Somehow related is also the question of the role of the cosmological constant (if the acceleration of the Universe is indeed due to a true cosmological constant) in the remote past of the contracting phase. When going backward in time, the universe inevitably becomes $\Lambda-$dominated at some point. This might be used to explain the matter content of the Universe by a purely geometrical origin and even lead to a cyclic scenario which does not suffer the problem of growth of inhomogeneities during the contracting phase \cite{Barrau:2014kza}.\\

In the case of an Euclidean phase, things are more complicated. Does it makes sense to ``propagate" perturbations in the absence of time? This is  possible if one interprets the effect as an instability in the equations of motion but the process is hard to understand if one really considers that there exists a true change of signature in the space-time structure. In fact, the naive propagation of perturbations across the bounce leads to a spectrum inconsistent with data \cite{Bolliet:2015raa}. This could be due to one of the numerous assumptions (isotropy, no backreaction, etc.) but it could also be an indication that the idea of propagating modes through an Euclidean phase does not make sense at all. An interesting alternative way of dealing with the same situation was suggested recently \cite{Bojowald:2015gra}, taking advantage of the known mathematical treatment of the Tricomi problem. Conceptually it opens a whole new perspective for cosmology: the mixed-type partial differential equations for modes in this context lead to a nice balance between deterministic cyclic models and singular big bang models. There is no initial divergence, and yet initial data in the infinite past do not uniquely determine the entire space-time structure. For every mode, it is necessary to specify one function at the beginning of the expanding phase even if initial values for the contracting phase had  already been chosen. Still, the normal derivative of the field is not free and may carry subtle but interesting information about the pre-big bang epoch.\\

A recent work suggested that time could emerge from a ``SO(4) $\rightarrow$ SO(1,3)" symmetry breaking \cite{Mielczarek:2012pf,Mielczarek:2012tn}. By analogy with solid state physics, one could speculate that this transition, exhibited by the deformed algebra approach and other LQC studies mentioned before, is a result of the symmetry breaking at the level of the fundamental structure of spacetime. In particular, one could assume that the original SO(4) spacetime symmetry is broken into SO(3), where the residual SO(3) is the rotational symmetry of triads. The time direction could therefore be seen as the order parameter of the symmetry broken phase.\\

Another important issue is related to the cosmological shear. As the shear term is proportional to $1/a^6$ in the contracting branch, if a causal evolution viewpoint is adopted, anisotropies become significant at the bounce, when the scale factor reaches its minimal value. The effect of anisotropies on the duration of inflation was studied \cite{Linsefors:2014tna, Gupt:2013swa} and it was shown \cite{Linsefors:2013bua} that if initial conditions are set at the bounce, there are many more solutions leading to a universe that does not resemble ours at all than to a universe with a standard classical limit. If initial conditions are set in the remote past of the contracting branch, the problem is automatically evaded by selecting a solution with the correct asymptotic behavior. Then, what should be the ``initial" conditions for the shear? This is a  delicate and important question. When dealing, for example, with the phase of the oscillations of the scalar field, a flat distribution can easily be chosen, especially because it is conserved in time. But there is no straightforward choice when the shear term is included, and the predictive power of the whole LQC approach depends crucially on this as the number of e-folds is strongly dependent on the anisotropies at the bounce.\\

In our opinion, the most important conceptual and technical issue is related to Planck length effects. In the black hole sector, it was recently suggested \cite{Rovelli:2014cta,Haggard:2014rza} that too much emphasis was put on Planck length effects, neglecting Planck density effects. The situation is somehow reversed in cosmology. As explained in the first section of this article, the background dynamics at high density seems to be under a fairly good control. This is the main and less controversial result of LQC. However, when calculating the primordial spectra, one faces a Planck {\it length} problem. One possibility, used by the authors of the dressed metric approach, is to fine-tune initial conditions so that the amount of inflation just equals the required minimum. In that case, modes of physical interest are never trans-Planckian. However, for the vast majority of the parameter space (including the value favored by the analysis with initial conditions in the past \cite{bl}) modes of physical interest are much smaller than the Planck length at the bounce time. This problem already exists in standard cosmology and is well known. However, it becomes much more important in the framework of a theory that predicts that there is nothing smaller length than the Planck length (This statement is rigorous about surfaces only, it is actually much less clear for lengths and depends on the chosen operator.) Otherwise stated, the question is: in the purely classical (far from the bounce) contracting branch of the Universe, what happens to a photon blue-shifted to the Planck length? It is possible that the dressed metric approach automatically accounts for such effects through its quantum field on a quantum background treatment. But the physical interpretation of values of wavenumbers higher that the Planck scale --and they are 30 orders of magnitude higher in a typical case-- is still to be understood. In the deformed algebra approach, this point is clearly not taken into account at this stage. It probably could be accounted for by modified dispersion relations. This makes sense as the LQC deformation of the GR algebra naturally leads to such effects. This will however in general raise new conceptual issues as complex frequencies would then enter the game.\\

Finally, we would like to mention that the problems of quantum-to-classical transition (the measurement problem) and entropy production are rarely addressed in the framework of LQC. They should be faced in future studies.\\

The issues presented in this article are in no way suggesting that LQC fails as an effective quantum cosmological formalism. In fact, the abundance of questions being addressed is a positive sign showing that LQC is an active and healthy field of research, with motivating challenges for the years to come.
\bibliographystyle{ws-ijmpd}
\bibliography{refsBoris}
 \end{document}